# Retrieval-Augmented Generation in Industry: An Interview Study on Use Cases, Requirements, Challenges, and Evaluation

Lorenz Brehme[1][a], Benedikt Dornauer[1][b], Thomas Ströhle[1][c],
Maximilian Ehrhart[2][d], Ruth Breu [1][e]

[1]*University of Innsbruck, Innsbruck, Austria*
[2]*CASABLANCA hotelsoftware, Schönwies, Austria*
{*lorenz.brehme, benedikt.dornauer, thomas.stroehle, ruth.breu*}@uibk.ac.at, maximilian.ehrhart@casablanca.at

Abstract: Retrieval-Augmented Generation (RAG) is a well-established and rapidly evolving field within AI that enhances the outputs of large language models by integrating relevant information retrieved from external knowledge sources. While industry adoption of RAG is now beginning, there is a significant lack of research on its practical application in industrial contexts. To address this gap, we conducted a semi-structured interview study with 13 industry practitioners to explore the current state of RAG adoption in real-world settings. Our study investigates how companies apply RAG in practice, providing (1) an overview of industry use cases, (2) a consolidated list of system requirements, (3) key challenges and lessons learned from practical experiences, and (4) an analysis of current industry evaluation methods. Our main findings show that current RAG applications are mostly limited to domain-specific QA tasks, with systems still in prototype stages; industry requirements focus primarily on data protection, security, and quality, while issues such as ethics, bias, and scalability receive less attention; data preprocessing remains a key challenge, and system evaluation is predominantly conducted by humans rather than automated methods.

*This preprint was accepted for presentation at the 17th International Conference on Knowledge Discovery and Information Retrieval (KDIR25).*

## 1 Introduction

Since 2022, Large Language Models (LLMs) have made significant progress in research and have gained widespread popularity [Google Trends, 2025]. However, they still face substantial challenges, such as hallucinations caused by insufficient context or outdated context, limited access to up-to-date domain-specific knowledge, and difficulties in verifying the accuracy of generated information [Zhang et al., 2023c, Wang et al., 2024]. To address these limitations, the concept of Retrieval-Augmented Generation (RAG) was introduced by [Lewis et al., 2021]. RAG extends LLMs by integrating external knowledge sources, enabling them to handle for example domain specific question answering (QA) task. By incorporating domain-specific information, RAG systems can generate more accurate, relevant, and contextually appropriate responses for specialized topics, and several such systems have already demonstrated the effectiveness of this approach [Gao et al., 2024]. Since the rise in popularity of RAG systems, organizations have been trying to build them for industrial use, often relying on large amounts of confidential and proprietary knowledge that standard LLMs do not contain [Zhou et al., 2025]. By leveraging corporate knowledge, RAG systems enable companies to develop tailored use cases and perform specialized, domain-intensive tasks that meet their unique needs.

However, the literature primarily focuses on the design and development of RAG systems themselves [Gao et al., 2024], with comparatively little attention given to their specific applications in corporate environments [Arslan et al., 2024]. This study aims to bridge this gap by examining the practical use of RAG systems in industry through an investigation of four key aspects—(1) their use cases, (2) the requirements set by companies, (3) the technical and organizational challenges encountered, and (4) the methods practitioners use to evaluate their quality—resulting in four research questions.

[a]  https://orcid.org/0009-0009-4711-2564
[b]  https://orcid.org/0000-0002-7713-4686
[c]  https://orcid.org/0000-0002-1954-6412
[d]  https://orcid.org/0000-0002-9554-0231
[e]  https://orcid.org/0000-0001-7093-4341

Building on [Arslan et al., 2024] six-category classification of industrial RAG applications, we investigated through our interviews which use cases are most prevalent and how they align with this framework:

> **RQ1:** How have companies applied RAG in their use cases and what opportunities do companies see when using RAG systems?

To identify key factors for the practical deployment of RAG systems, we conducted a literature review and analyzed how companies address them, resulting in 12 industry-relevant requirements (e.g., security and data protection). Our goal was to capture practical, real-world perspectives rather than purely theoretical ones by investigating:

> **RQ2:** What requirements do companies set for RAG systems, and how are these requirements implemented in practice?

Furthermore, we aim to identify common challenges companies face when implementing RAG systems, as repeated mistakes can lead to wasted resources. By understanding these challenges and how they are addressed, we can help inform more effective adoption strategies and implementation guidelines. Following, our interviews also explored lessons learned during implementation and whether industry practices reflect improvements suggested in the literature, such as those outlined by [Zhao et al., 2024]. Based on these insights, we provide practical recommendations on key priorities and common pitfalls to avoid when adopting RAG systems.

> **RQ3:** What challenges do companies face when implementing and using RAG systems, and what lessons have been learned from these experiences?

Lastly, we investigate how industry practitioners assess the quality of their RAG systems, focusing on whether and how they have adopted evaluation methods from academic research in their practical applications. Academic research has proposed several automated evaluation methods for assessing entire RAG systems [Brehme et al., 2025b]. In this study, we examine whether and how these methods have been adopted in industry by exploring if practitioners evaluate their systems against the specific requirements they identified and measure the extent to which these requirements are fulfilled:

> **RQ4:** How do industry practitioners evaluate the quality of RAG systems?

To explore these questions, we conducted 13 semi-structured interviews, following stepwise guidelines [Adams, 2015] and a systematic five-step qualitative procedure [Schmidt, 2004], focusing on industry experts directly involved in the implementation of RAG systems. Our main findings are as follows:

- The most common applications of implemented RAG systems are QA tasks, with each RAG typically designed for a specific use case within a limited domain.
- The technical readiness level in most companies remains at the prototype stage.
- The industry's main requirement focus is on data protection, security, and quality, whereas most of the time ethical considerations, bias mitigation, costs, and scalability are less prioritized.
- One of the main challenges is data preprocessing for preparing the RAG system, which is critical for the overall system quality.
- RAG systems are primarily evaluated by humans, rather than through automated methods using LLMs.

## 2 Related Work

This section explores three key aspects: types of RAG applications, various enhancements to improve RAG systems, and approaches for evaluating their performance.

**RAG applications.** In [Arslan et al., 2024], six main types of RAG applications in industry are identified through a comprehensive review. (*i*) The most common is *question answering*, with examples such as medical QA tasks [Xiong et al., 2024] or commonsense QA [Sha et al., 2023]. (*ii*) RAG is also widely used for *text generation and summarization*, for instance, generating stories with complex plots [Wen et al., 2023]. (*iii*) *Information retrieval and extraction* represents another key area, supporting tasks like regulatory compliance

QA in the pharmaceutical industry [Kim et al., 2025]. (*iv*) *Text analysis and preprocessing*, such as sentiment classification [Mahboub et al., 2024], further illustrate how RAG systems support data understanding. (*v*) In addition, RAG contributes to *software development and maintenance*, for example through code completion [Lu et al., 2022]. (*vi*) Finally, *decision making and applications*—such as automated cash transaction booking [Zhang et al., 2023a]—demonstrate how RAG can assist in operational processes.

**Enhancements to improve RAG systems.** Another aspect of RAG is system enhancement, as highlighted by [Zhao et al., 2024], who identify several RAG improvements in the literature, including input enhancements as a key category. One type of input enhancement is query transformation, which involves approaches such as refining or rewriting the input query to improve retrieval effectiveness [Chan et al., 2024, Tayal and Tyagi, 2024], while another is data augmentation, wherein semantic-preserving transformations are applied to enhance the database, as demonstrated in the pre-training of code retrieval models [Lu et al., 2022]. The second category of enhancements centers on retriever improvements, including techniques such as recursive retrieval [Yao et al., 2023], hybrid retrieval [Yu et al., 2022], re-ranking methods [Glass et al., 2022], and chunk optimization strategies [Sarthi et al., 2024]. The third category is generator enhancements, which includes techniques like prompt engineering [Wei et al., 2023] and fine-tuning of the decoder [Jin et al., 2023] to improve the performance of the generation component. The fourth category is result enhancement, which involves techniques for rewriting or refining the generated output [Liu et al., 2024]. Finally, the fifth category is RAG pipeline enhancement, which refers to the development of new or optimized RAG pipeline architectures [Zhang et al., 2023b].

**Evaluating RAG systems.** Quality is a critical aspect of RAG systems, with several evaluation frameworks designed to measure various quality requirements. One such framework is RAGAS [Es et al., 2023], which assesses retriever quality by evaluating the relevance of retrieved documents. This can be done using an LLM [Es et al., 2023], an already prepared test set with labeled datasets [Tang and Yang, 2024], or human judgment [Afzal et al., 2024]. Another important aspect is the generator quality, where the generated answer is evaluated for quality attributes like faithfulness [Es et al., 2023]. In this case, LLMs [Es et al., 2023], embedding techniques [Kukreja et al., 2024], token-based approaches [Li et al., 2024], or human evaluators [Pipitone and Alami, 2024] may be employed. Additionally, the overall performance of the system is measured by factors such as retrieval time and processing speed [Kukreja et al., 2024]. These various quality metrics help assess the effectiveness of each component of the RAG system and ensure optimal performance.

Most existing research focuses on broad use cases, offering little insight into how RAG systems are applied in real-world corporate settings. Furthermore, there is a noticeable lack of empirical qualitative studies exploring the challenges organizations face when adopting and utilizing these systems. To address this gap, our research focuses on industrial use cases, enhancement strategies, and evaluation approaches.

## 3 Methodology

For conducting our semi-structured interviews, we stuck to the step-wise guidelines by [Adams, 2015], considering in addition the recommendations for software engineering interviews by [Hove and Anda, 2005]. Each interview was scheduled to take approximately one hour and was structured into five main parts: Company and Interviewee Background, Adoption and Implementation of RAGs, Requirements for RAGs, Quality Assessment of RAGs, and Outcomes and Future Outlook, with mostly open questions and few closed questions, as recommended by [Adams, 2015]. Before conducting the semi-structured interviews, the interview guidelines were pilot-tested with one individual to refine the guide.

We selected purposive sampling [Campbell et al., 2020] to contact companies from the authors' network that apply LLMs. For selecting respondents and arranging interviews, we specifically request in our email that experts who have been directly involved in implementing a RAG system be identified. Furthermore, along with our request, we attached our interview questionnaire. We selected participants from diverse company sizes, domains, and experience levels to ensure a heterogeneous sample and gain domain-independent insights.

The first author asked the questions following the questionnaire, while another author took notes and posed follow-up questions when responses were incomplete or when further clarification was needed. Additionally, each interview was fully recorded via Microsoft Teams, transcribed, and subsequently made available for detailed analysis. In total, 13 interviews were conducted between early April and early June 2025. Table 1 lists the main key facts of the individual participants, showing the wide variety of domains represented.

Based on the gathered materials, we simultaneously started with the analysis of the interviews, using a five-step procedure described by [Schmidt, 2004].

1. Firstly, we filter, separate, and summarize the interviews in a table-like to get a general overview and find similarities and distinctions.

2. Next, we draft analytical categories compiling them into a coding guide with definitions and variants, leading to the four RQs.

3. We then code the interview using predefined categories to enable comparison, making adjustments through team consensus when necessary.

4. After the coding, we present the total overview and mapping, in most cases, in a table-like way.

5. Finally, we conduct an in-depth analysis of related cases through repeated readings, knowledge extraction, and textual synthesis, which are then integrated into the written paper.

In addition, to answer RQ2, we first performed a rapid literature review to derive implementation-relevant requirements for RAG systems. Subsequently, we used descriptive statistics to evaluate how these 12 predefined requirements were assessed during our interview study. Therefore, the participants were asked to rate the 12 predefined requirements on a scale from 1 to 10 based on their importance. This provided an overview of where the actual focus lies and helped us understand which requirements are considered important, which are seen as difficult to achieve, and why some may be intentionally prioritized.

More details and the complete replication package are provided on our Zenodo repository [Brehme et al., 2025a].

## 4 Findings

### 4.1 RQ1: Use Case and Goals

We collected use cases through interviews and, based on our coding, identified similarities with those proposed by [Arslan et al., 2024]. As a result, we adapted their categorization, which is presented in Table 2.

---

[2]Company Size based on OECD [OECD, 2017]: Small Comp. [employees≤49] (SC), Medium Comp. [50≤employees≤249] (MC), Large Comp. [employees≥250] (LC).

Table 1: List of interview participants with IDs assigned alphabetically and listed chronologically.

| ID | Size[2] | Specific Domain | Research start |
|---|---|---|---|
| A | MC | Hotel Software | 2024 |
| B | LC | Medical | 2018 |
| C | SC | Software Development | 2023 |
| D | LC | Banking Software | 2023 |
| E | LC | Multi Engineering | 2021 |
| F | SC | Consulting | 2022 |
| G | MC | PI Business Consulting | 2024 |
| H | LC | Media | 2023 |
| I | LC | Software for Municipalities | 2023 |
| J | SC | AI | 2023 |
| K | MC | Process Automation | 2022 |
| L | MC | Web Application | 2023 |
| M | LC | Software Consulting | 2022 |

In most cases, RAG is employed to facilitate knowledge transfer by delivering information in a human-readable format through a conversational chat interface [A,B,D,E,F,G,H,I,J,L,M]. According to the categorization by [Arslan et al., 2024], this application falls under the category of *question answering*. These RAG systems enable users to perform targeted searches across various departments within an organization to address specific questions. For instance, one participant described using RAG to retrieve information from HR documents [D], while others accessed data from ticketing systems [F,I]. Overall, these RAG systems are primarily employed to process internal documents and extract relevant information in order to provide precise answers to user questions.

Other mentioned use cases fall under the *Decision Making and Application* category: RAG is being extended to support operational tasks, going beyond information retrieval to perform actions such as creating support tickets or generating reports [C]. The goal is to combine the retrieval of relevant information with the execution of operational tasks. Leveraging access to internal wikis, ticketing systems, and source code repositories, the RAG system can, for example, create new tickets, initiate projects, or generate structured reports based on the retrieved information. Furthermore, RAG systems are also being adopted to replace legacy systems [M]. Last but not least, one use case involves document formatting. In this context, documents are provided as examples of the desired format, and the LLM is expected to generate outputs that adhered to the same structure and style. This functionality was integrated into a larger application and represented one step within a broader automation workflow [K]. The last use case fits in the *Text generation and Summarization* category of [Arslan et al., 2024].

Table 2: List of use cases by interview participants

| Category | ID | Primary Use Case |
|---|---|---|
| Question Answering | A | Search Assistant for Support |
| | B | Medical QA |
| | D | Chat Assistant with Specific Expertise |
| | E | Find Information from Repository |
| | F | Search Assistant for Different Tools |
| | G | Modular RAG System for Companies |
| | H | Get Information About Other Employees/Team Group Messages |
| | I | QA for Ticketing System |
| | J | Automatic Customer Support |
| | L | Search Documentation |
| Decision Making and Application | C | Chatbot for Customer with ChatOps |
| | M | Automation of Legacy Systems |
| Text Generation and Summarization | K | Contract Formatting |

Another important aspect of the identified use cases is understanding the underlying goals for adopting RAG systems. The first perspective that emerged was the rationale for choosing RAG over other technologies, such as standalone LLMs. One key reason was the ability to easily **update and expand the knowledge** base with company-specific information [C,E,L]. This offered greater flexibility compared to fine-tuning a model, which is more static and resource-intensive. With RAG, the underlying LLM could also be swapped out for a more advanced model as they become available, without the need to retrain from scratch [E]. Participants also noted that RAG can **reduce hallucinations** and generally improve the quality and reliability of the output by grounding responses in retrieved, contextually relevant documents [C,H,J,K,L]. The most frequently mentioned goal was to **save time and increase efficiency** in daily work processes [A,D,F,G,I,M]. This objective is closely linked to **reducing the workload** on human resources [B,I]. Participants explained that by using RAG, less time was spent on manual information retrieval, allowing employees to focus on more value-added tasks. An additional advantage of RAG is the **easy access to relevant information** without the need for complex search queries [A,C,H,I], especially when it is fragmented across various departments or held by multiple individuals [A]. Participants emphasized the importance of centralizing and structuring this dispersed knowledge to ensure effective use. Additionally, RAG was seen as a way to reduce knowledge dependency on specific individuals—mitigating the risk of knowledge loss when key personnel leave the company [I]. This points to broader opportunities, such as enabling company growth without a proportional increase in human resources, due to improved efficiency and automation [B]. Furthermore, participants noted that accelerating the information retrieval process helps **employees maintain focus**. Instead of being interrupted or distracted by time-consuming documentation searches, they can stay engaged with their core tasks [D].

A final consideration is the future development of RAG systems and their expanding capabilities. One emerging direction is the evolution toward agentic RAGs [A,D,G,M], where these systems acquire decision-making abilities and operate autonomously within defined workflows. This concept is already being explored in research [Singh et al., 2025a] and is likely to be adopted by industry in the near future. Another significant advancement is the shift toward multimodal RAGs [G], which can process and analyze diverse input types—such as text, images, and structured data—rather than being limited to a single document format. This development will further enhance RAG pipelines, enabling more versatile and powerful applications.

## 4.2 RQ2: Requirements

For effective implementation, RAG systems require a well-defined set of infrastructural, technical, and organizational prerequisites. Based on our literature review, we set up a list of requirements that we aimed to validate in terms of their importance for RAG systems in industrial settings. The results of the importance scores is presented in Figure 1.

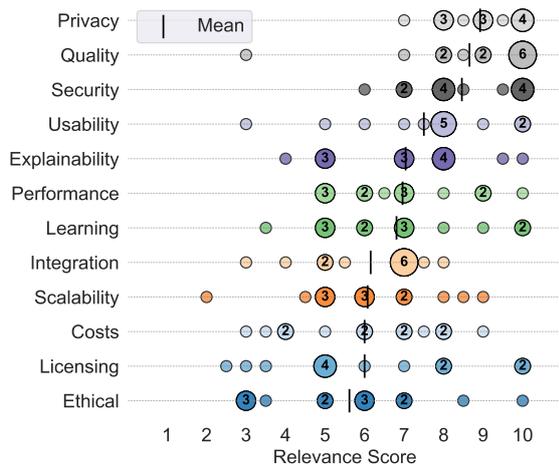

Figure 1: Sorted distribution of requirement relevance scores, averaged by mean. Scores range from 0 (not relevant) to 10 (highly relevant).

**Security and Data Protection** in RAG systems means to ensuring that the system cannot be compromised or "jailbroken," and that it does not disclose data outside its intended boundaries. This includes anonymizing any confidential data and safeguarding user information against misuse. Initially, these aspects were treated as separate requirements, but based on interview feedback, we merged them into a single, interdependent requirement because they are closely related and must be addressed together. The importance of security and data protection varied depending on the use case. For internal systems accessible only by employees, this requirement was considered less critical but still relevant. However, it becomes significantly more important when the system is accessible externally [D]. Additionally, the emphasis on this requirement differed by company size and geographic region. In large companies and in Europe, where data protection regulations are strict, security was rated as highly important [I]. In contrast, one participant mentioned that in the US and Asia data protection is generally less prioritized in their contexts [J]. Overall, security and data protection were regarded as a very important requirement across most cases.

**RAG Quality** denotes the overall quality of the responses generated by the RAG system. This includes the quality of the underlying data, the retriever component, and the generated answers. Initially, we identified RAG quality and answer quality as separate requirements, but we extended this to include data quality after it was explicitly mentioned by participants [C,D,H,I,J,M]. The intent behind this requirement is to ensure that the RAG consistently delivers high-quality, relevant, and accurate responses. All participants rated this requirement as highly relevant, with scores of 8 or higher, except for one [K], who rated it a 3. In that specific use case, RAG was only a small component within a larger process, and low-quality outputs could simply be discarded without affecting the overall system.

**Usability** is defined as how easy and intuitive the RAG system is for users to interact with. The perceived importance of usability varied among participants. Some stated that usability was not a major concern because users were already familiar with similar chat interfaces, for example ChatGPT or Gemini, making adaptation straightforward and requiring minimal effort [A,E]. Others emphasized that usability is crucial, as users are more likely to adopt the system if it is intuitive and easy to use [C,G]. Overall, we conclude that usability is important but generally easy to achieve, given that conversational interfaces are now common [C,D].

**Explainability/Transparency** refers to the extent to which the RAG system provides clear and understandable explanations for its answers. This is closely tied to the configuration of the RAG, where the system must present the sources of information transparently. Specifically, the retrieved documents supporting each part of the answer should be visible, allowing users to understand why the system generated a particular response based on these documents. Nine participants rated this requirement as relevant, assigning it a score of seven or higher.

**Performance** covers aspects related to system speed and responsiveness. This requirement was generally not considered highly important by the participants. For example, Interviewee F mentioned that there are strategies to manage slower system responses, such as employing "thinking" delays that give the impression of a more thoughtful output. Overall, performance was not emphasized as a critical requirement, and most participants rated its relevance as low.

**Continuous Learning** describes the ability of the RAG system to update and expand its knowledge base to stay current with new information. Initially, we grouped this requirement together with continuous operation, but during analysis, we identified them as distinct aspects. Continuous learning was considered particularly important in use cases where data freshness is critical.

**Continuous Operation** encompasses the ability to maintain, monitor, and update the RAG system during deployment. For example, it was highlighted the importance of benchmarking the system regularly, especially after changes in data or system components [G,I,H]. Continuous operation includes tasks such as updating components or adjusting configurations while the system is live. While this requirement received moderate ratings, it was generally overshadowed by the higher importance placed on continuous learning. Therefore, continuous operation remains relevant but is considered of moderate priority compared to continuous learning.

**Integration in Setup** describes the requirement that the RAG system can be seamlessly integrated into the existing technical infrastructure. For example, the company has contracts with a service provider, already uses its services, and prefers to remain within this environment. This includes ensuring compatibility with current databases, cloud services, and other components already in use. The importance of this requirement varied depending on the company's context. For one participant [F], integration was less critical because they were able to build the RAG system from scratch without relying on existing services. Conversely, in larger organizations with established contracts and infrastructures, adapting the RAG system to fit into existing technologies was necessary and therefore more important [H]. Overall, this requirement was rated as moderately relevant, with even those who considered it important assigning relatively low scores.

**Scalability** addresses two dimensions that we consider: scaling the number of users and inputs, and scaling the knowledge base of the RAG system. Some participants generally believed that scaling is straightforward since most LLMs are accessed via APIs that inherently support scaling [A,F]. Similarly, cloud-based data storage solutions make scaling the knowledge base manageable without significant challenges. One participant mentioned that this requirement is seen as less important because some companies are still in the pilot phase and intend to address scaling at a later stage [E].

**Costs** encompass all expenses related to both the development and operation of the RAG system. This requirement was considered very relevant by some participants [C,H,I,M], who gave it ratings above eight. On the other hand, it was rated as less relevant by others, for various reasons. Some participants noted that the cost of LLM usage is expected to decrease over time, reducing financial concerns [K,B]. Additionally, since most organizations are still in the research and development phase, cost considerations were not yet a significant challenge. Instead, the focus remains on determining the feasibility and practical implementation of RAG systems. Interviewees anticipated that cost will become a more critical factor in later stages of RAG deployment [A]. In general, it is difficult to demonstrate that the costs are justified and that improvements in efficiency are both measurable and meaningful [E].

**Licensing and Copyright** concerns whether the output generated by the RAG system can be used freely without infringing on copyright-protected content. This issue primarily arises from the underlying LLMs, where copyright ownership and usage rights are not always clearly defined. Some participants expressed concerns about potential legal challenges when using generated content [H,M]. However, participant B noted that such issues are increasingly being addressed, for example, through regulations like the European AI Act, which clarifies copyright ownership of AI-generated content. Overall, this requirement was not a major focus for most participants, with many either unaware of it or not considering it a pressing concern.

**Ethical Considerations and Bias** require that the RAG system produces neutral, unbiased responses and avoids discrimination or unethical outputs. Despite the importance of ethics in AI, this requirement received the lowest relevance ratings from participants. Ethical considerations and biases were either not considered relevant to their use case [F] or had not reflected on the issue at all [J,M]. Overall, ethical considerations and bias mitigation were not seen as pressing issues at the current stage of RAG development. However, this largely depended on the company: some regarded these aspects as relevant [B,C] with a rating above 8, while others did not [D,G,K] with a rating of 3. The relevance ratings varied widely in this regard.

## 4.3 RQ3: Identified Challenges and Lessons Learned

First introduced in 2021 by [Lewis et al., 2021], RAG has since gained increasing popularity, as reflected in the number of companies adopting it over time (e.g., based on our interviews: 1 company in 2021, 3 in 2022, 6 in 2023, and 2 in 2024). Since their start, they encountered various challenges that have been subsequently summarized and categorized into four categories: the data management, the retrieval process, the generator component, and those related to the overall process, represented in Figure 2.

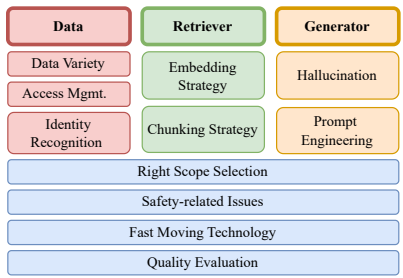

Figure 2: Overview Industry Challenges

**Data:** One main issue is the **data variety** with the unstructured nature of the data, which is often dispersed across various systems and exists in different formats, such as PDFs, images, and other document types. Pre-processing this data and extracting relevant information requires substantial effort [A,E,I,M]. Another practical challenge identified was **access management**. Not all users should have access to all data, and it is crucial to ensure that data from one project does not overlap with another. Interviewees highlighted that managing data access and maintaining project-specific data boundaries were significant challenges during implementation [D,E,F,H,J].

Another key issue emphasized by participants is **identity recognition**. Specifically, it can be challenging to determine whether two entries refer to the same entity or simply share a similar name [I,H]. Participant H recommended implementing a unified identity management system across the organization. Such a system simplifies the process of identifying and linking entities, thereby reducing confusion and minimizing errors.

This problem is further complicated by the use of abbreviations [I]. In many cases, the same abbreviation can represent different concepts across documents, making it challenging to ensure that the RAG system retrieves the correct meaning in each context. To address conflicts in identity recognition and abbreviations, participant I opted to use ontologies within a knowledge graph instead of a vector database. This approach helps ensure that entities are correctly linked, unambiguously referenced, and also supports the identification of synonyms or abbreviations. Moreover, certain types of data—such as addresses or product pages—are often highly similar, making it challenging to distinguish between them accurately [B,K] To solve these challenges, the critical importance of data quality was highlighted [A,M]. They recommended investing substantial time and effort in data preparation, as high-quality data leads to significantly better results.

**Retriever:** The second category of challenges is related to the retriever component of the RAG system. Determining the appropriate **chunking and embedding strategies** further complicates the process. B, E, and L explained that while a basic RAG system can be easily adopted with good initial results, working with real-world data is far more complex. In practice, selecting and collecting the right data and then choosing an optimal chunking and embedding approach proved to be a significant challenge. Even deciding on the best way to perform chunking was a problem. In terms of chunking strategies, they advised maintaining an optimal chunk size—not too small, to avoid fragmented context, and not too large, to prevent overwhelming the LLM with excessive information [B,C].

**Generator:** The last challenge pertains to the generator component, concrete to the issue of **hallucination** in LLMs. It was highlighted that hallucinations—instances where the model generates incorrect or misleading information—are a significant obstacle during implementation [A]. Specifically, the LLM may fail to accurately convey the information retrieved by the retriever component, or it may introduce erroneous details that were not part of the retrieved data. It also depended heavily on the user, as the results varied significantly when different prompts were used [C]. To address this challenge of inconsistent outcomes caused by varied user prompts or hallucination, a solution mentioned was to implement a RAG system where user input is automatically refined and improved by the LLM [A,C]. This approach minimizes the impact of poorly crafted prompts, ensuring more accurate and consistent results. Additionally, users should be made aware that RAG systems do not always produce correct answers. It is important to emphasize the need for critical evaluation of the output, as errors may occur and should always be taken into account [A] Ensuring that the RAG system consistently produces reliable answers was challenging, and evaluating this consistency proved equally difficult [I]. Another critical challenge is **prompt engineering**. It can be difficult to design the right prompt for a specific model to achieve optimal results [A,C,K]. A emphasized that even slight modifications to the system prompt can significantly disrupt the entire system's performance.

**Overlapping concerns:** One challenge, as highlighted by D, is determining the **right scope** for RAG implementation. Companies must decide whether to include all organizational data or focus on a limited, specialized subset. Defining the appropriate amount of knowledge and setting clear boundaries is a complex task. The next challenge is related to **safety issues** of the RAG systems. Participants raised concerns about potential risks, such as jailbreaking the model to bypass restrictions [M] or exploiting user interactions to collect employee data [H]. Such misuse can have serious privacy and security implications. Participants B and E expressed difficulties in keeping up with the **fast-moving technologies**. They found that by the time one method was implemented, it was often already outdated due to the emergence of new, more efficient approaches. This constant evolution creates uncertainty, as there is no established best practice for implementing and maintaining RAG systems. It also complicates efforts to address the previously mentioned challenges. The **quality evaluating** of the RAG system quality is also problematic. Unlike traditional software, where formal benchmarks can be applied, RAG performance is highly context-dependent, varying based on user input. C noted that different users adapt to the technology at different speeds.

**Implementation Recommendation and Lessons Learned** We can outline several recommendations for implementing a RAG system: The first and most critical step is identifying a suitable use case. This use case must have a clearly defined process, including a well-documented and justified automation goal [F,I,M]. Once the use case is established, the next step is data selection. The data must be tailored specifically to the chosen use case and should not be overloaded with excessive or irrelevant details [B,C]. It is essential to focus on a narrow and consistent domain, avoiding overlap with other departments or projects [D,F]. Ideally, a separate RAG implementation should be created for each project or department, with a clearly scoped and precise objective [D,F]. Following this, significant effort should be invested in data preparation and preprocessing to ensure high-quality inputs for the RAG system [A,M]. To support this, it is crucial to identify an effective chunking strategy. Chunks should be large enough to contain sufficient context for answering questions accurately, but not so large that they overwhelm the generator or exceed processing limits [B,C]. Striking the right balance is essential to ensure both retrieval relevance and generation quality. Regarding the generator component, it may be beneficial to integrate additional modules that modify or structure the prompt automatically. This can help mitigate the risk of poor prompting by end users and improve the overall quality of the generated responses [A,C].

As a general recommendation, the system should be designed with a modular architecture. Each component should be easily replaceable, allowing for the seamless integration of new LLMs or emerging technologies [F]. However, not every new advancement should be adopted immediately. If an existing system is stable and performs well, it is often more efficient to continue using it rather than investing significant time and resources into integrating newer solutions [B,E]. This also highlights that there is no „one-size-fits-all "solution when it comes to choosing an LLM or other components for RAG systems [B,E,H]. Instead, each RAG may require a different architecture, depending on the specific use case, data type, and desired outcomes.

Nonetheless, regular and systematic evaluation should be conducted, particularly after any changes to the system. If the underlying data or any parameter of a component is modified, a comprehensive re-evaluation is necessary. Continuous evaluation should also be part of the deployment phase to ensure the system maintains consistent performance in real-world conditions [B,I,K].

### 4.4 RQ4: Industrial RAG-Evaluation

In research, various evaluation frameworks have been proposed [Brehme et al., 2025b], ranging from fully automated to entirely manual approaches. In industrial use cases, evaluations were predominantly performed manually by human experts [A,B,C,E,I,J,L,M].

For this **manual approach**, testers analyzed the retrieved documents and generated answers to assess whether they are correct and relevant to the input queries, based on their human perception. This was sometimes further unified by selecting a group of testers who manually reviewed the RAG outputs using a predefined question test set [I,J]. In some cases, a user feedback UI mechanism was also integrated, like thumbs up/down buttons, allowing users to rate responses [A,B]. These ratings were then reviewed to identify issues and iteratively improve the system. In some cases, this was further extended by incorporating categories, such as tone preferences (e.g., „I do not like the tone.") to refine the evaluation process [B,H]. Apart from participants [I,B], who applied specific evaluation metrics, such as tone [B], correctness [I], coherence [I], answer quality [I], robustness [I], and the alignment between the retriever and generator components [B,I], the remaining participants primarily relied on their personal experience for assessment.

Only two participants described the use of **automated evaluation methods with the help of AI** [K,G]. For participant K, the system output was binary (i.e., correct or incorrect), enabling them a straightforward evaluation based on boolean values. The test dataset was constructed from anonymized historical queries, allowing for the evaluation of RAG within an existing workflow where only a system component had been replaced [K]. In interview G, an automated pipeline was implemented, utilizing an LLM to assess the quality of RAG-generated answers based on specific metrics, for instance, accuracy and tone (e.g., friendliness). Several other participants mentioned plans to implement automated testing using test datasets in the future [A,H].

Beyond response quality metrics, interaction rate exists as another frequently monitored indicator. Therefore, several participants reported tracking the number of daily requests to monitor how usage evolved over time, using this data as an indicator of user acceptance and adoption [B,C,H]. Additionally, they monitored the frequency and nature of issues encountered while interacting with the RAG system.

Performance-related metrics, such as latency and system responsiveness, were also used to assess the operational quality of the RAG implementation [B].

## 5 Discussion

RAG systems are gaining increasing attention in industry, building on extensive research conducted over the past year [Arslan et al., 2024].

Consistent with the use cases reported by [Arslan et al., 2024], the *Question Answering* category emerges as the most prominent application in industrial settings. Functionalities extending beyond traditional question answering in chatbots remain uncommon and were identified in only three instances, where, for instance, a participant incorporated operational capabilities into their RAG implementation. In contrast, [Arslan et al., 2024] outline a broader set of application scenarios, which may represent the next stage of development for industrial adoption, as many companies are still in the early stages of building RAG systems. This early-stage maturity was further reflected in our discussions about Technology Readiness Level (TRL): 12 out of 13 participants rated their use cases below TRL 7, suggesting that most RAG solutions are still in development or prototyping and have not yet reached large-scale deployment.

Moreover, our findings show that all RAG implementations are currently being developed for narrowly defined use cases, such as HR or customer support, rather than as general-purpose solutions. This may be due to the complexity of tailoring retrieval and generation processes to domain-specific data, as well as concerns around accuracy, compliance, and user trust. The results suggest that organizations are prioritizing controlled, high-impact applications where the benefits of RAG can be clearly measured and risks are more easily managed.

Considering the twelve key requirements across the thirteen use cases, the aspects rated as most critical were RAG/Answer Quality (average rating: 8.7), Security (8.5), and Privacy/Data Protection (8.9), while surprisingly, Ethical Considerations and Bias Awareness received comparatively low attention, with an average rating of only 5.6. From discussions with the companies, we observed that they initially focus on specific requirements to achieve early results, while also addressing other important, but less critical, requirements for success during later stages of the RAG system's development. They remain cautious about using LLMs due to concerns over potential hallucinations and misinformation, which could result in reputational damage. One possible reason mentioned for companies' caution is that current RAG systems do not yet fully comply with the stringent requirements of the EU AI Act [Almada and Petit, 2025].

It became clear that companies are investing heavily in the development of RAG systems, driven by high expectations and strong confidence in this emerging technology: The primary motivation for companies to pilot and deploy such RAG systems is the expectation that they will enhance employee productivity by minimizing the time and effort spent on labor-intensive information retrieval tasks. This anticipated benefit is supported by a McKinsey report [Chui et al., 2012], which estimates that employees dedicate approximately 28% of their average workweek to searching for and gathering information.

We identified several challenges faced by companies during the current phases of RAG implementation, along with a set of frequently mentioned technical recommendations. Existing research offers a wide range of RAG enhancement strategies [Zhao et al., 2024], which are also being adopted in industry practices. For example, enhancements in query transformation and critical techniques in data preparation, such as chunk optimization, have been adopted in industrial settings. Also, data preprocessing remains a significant challenge in developing RAG systems, though AI-driven tools and processes may offer valuable support in automating and optimizing these tasks [Sivathapandi, 2022].

However, some promising recent research findings have not yet been implemented in practice. Participants mentioned these only as part of future plans. For instance, this includes the implementation of agentic RAG systems [Singh et al., 2025b] or personalized RAGs for use over private documents [Ryan et al., 2025]. Another aspect concerns the automatic assessment of RAG quality, as exemplified by RAGAS [Singh et al., 2025b]. Surprisingly, our findings indicate that RAG evaluation in industry is predominantly performed manually, without leveraging AI-based automation. This may be attributed to the lack of domain-specific test datasets, which are expensive and labor-intensive to collect and can quickly become outdated. This discrepancy highlights again a gap between academic research, which has developed multiple automated workflows for testing RAG systems [Brehme et al., 2025b], and current industry practices.

## 6 Conclusion

This study investigated the current state of RAG in industry through semi-structured interviews with 13 practitioners. The findings reveal a notable gap between academic advancements and industrial adoption. Industry use cases remain in the early stages of exploration, with most efforts limited to trials or pilot projects, and only a few achieving full deployment. Based on practitioners' experiences, we identified a comprehensive list of requirements critical for RAG implementation, highlighted cross-company challenges in adopting RAG, and summarized key lessons learned from practical deployments. Furthermore, the study emphasized approaches to RAG evaluation.

The limited number of interviews, predominantly from Europe, and the fact that the results reflect a specific point in time may limit the generalizability of the findings. Nonetheless, we think the consistency of responses suggests that the principal industry perspectives were effectively captured.

Looking forward, despite the ongoing trial phase, we expect industry to increasingly adopt recent advancements in RAG research, as the benefits outweigh implementation efforts. In particular, we expect that agentic RAG approaches are poised for widespread adoption due to their potential to enable greater system autonomy, dynamic adaptation to contextual changes, and seamless integration with external tools and services—capabilities that are currently lacking in industry applications.

## ACKNOWLEDGEMENTS

We would like to thank all our study participants, including Vaadin, Akkodis, cccom Moser GmbH, roosi GmbH, PPI AG, Flex.Insight Advisory, Kufgem GmbH, HerculesAI, DeepOpinion and other partners, for their valuable contributions. This research was supported by the Austrian Research Promotion Agency (FFG) under the GENIUS project [931318, 921454]. During the preparation of this paper, we used ChatGPT-4.1 as well as Grammarly for grammar and spelling checks, and GitHub CoPilot.